\pdfoutput=1
\documentclass[conference]{IEEEtran}
\usepackage{caption}
\usepackage{amsmath}
\usepackage{amssymb}
\usepackage{graphicx}
\usepackage{enumitem}
\usepackage{algorithm2e}
\usepackage[noadjust]{cite}
\makeatletter
\newcommand{\removelatexerror}{\let\@latex@error\@gobble}
\newcommand{\etal}{\textit{et al.}}
\usepackage{multirow}
\IEEEsettopmargin{t}{0.80in}
\begin{document}

\title{A Practical Searchable Symmetric Encryption Scheme for Smart Grid Data}

\author{
    \IEEEauthorblockN{Jiangnan Li, Xiangyu Niu, Jinyuan Stella Sun}
    \IEEEauthorblockA{Department of Electrical Engineering and Computer Science
    \\University of Tennessee, Knoxville
    \\\{jli103, xniu\}@vols.utk.edu, jysun@utk.edu}
}

\maketitle

\begin{abstract}

Outsourcing data storage to the remote cloud can be an economical solution to enhance data management in the smart grid ecosystem. To protect the privacy of data, the utility company may choose to encrypt the data before uploading them to the cloud. However, while encryption provides confidentiality to data, it also sacrifices the data owners' ability to query a special segment in their data. Searchable symmetric encryption is a technology that enables users to store documents in ciphertext form while keeping the functionality to search keywords in the documents. However, most state-of-the-art SSE algorithms are only focusing on general document storage, which may become unsuitable for smart grid applications. In this paper, we propose a simple, practical SSE scheme that aims to protect the privacy of data generated in smart grid. Our scheme achieves high space complexity with small information disclosure that was acceptable for practical smart grid application. We also implement a prototype over the statistical data of advanced meter infrastructure to show the effectiveness of our approach. 
\end{abstract}

\begin{IEEEkeywords}
searchable symmetric encryption, smart grid, privacy preserving
\end{IEEEkeywords}

\section{Introduction} \label{sec:intro}

The legacy power grid is in its way to become smart with diverse data-driven approaches. Smart grid, which introduces advancement in sensing, monitoring, control and communication to legacy grids, is considered to be the next generation power grid that can provide high-quality service to the public \cite{b1}. Moreover, as one of the most critical infrastructures in modern society, power grids are also complex and consist of enormous components. A well-known conceptual model of the smart grid is proposed by the U.S National Institute of Standards and Technology (NIST) in 2010 \cite{b2}. The model separates the smart grid into seven domains, namely generation, transmission, distribution, customers, markets, operations, and service provider. All these subsystems and components can become the sources of the large volume miscellaneous data in future smart grid systems. Some data sources in the smart grid are shown in Fig. 1. 

However, due to the divergence in structure, type and generation rate, how to integrate, store, and manage the data is still one of the active research fields in the smart grid community. Recently, research on remote cloud-based storage and management of smart grid data is becoming popular \cite{b3}. Arenas-Martinez \etal \cite{b4} presented and compared a series of cloud-based architectures to store and process smart meter reading data. Based on the specific characteristics of smart grid data, Rusitschka \etal \cite{b5} proposed a cloud computing model of ubiquitous data storage and access. In fact, outsourcing the data storage to the cloud can be an effective solution and has advantages in scalability, performance, and interoperability. 

\begin{figure}[htbp]
\centerline{\includegraphics[width=1\linewidth]{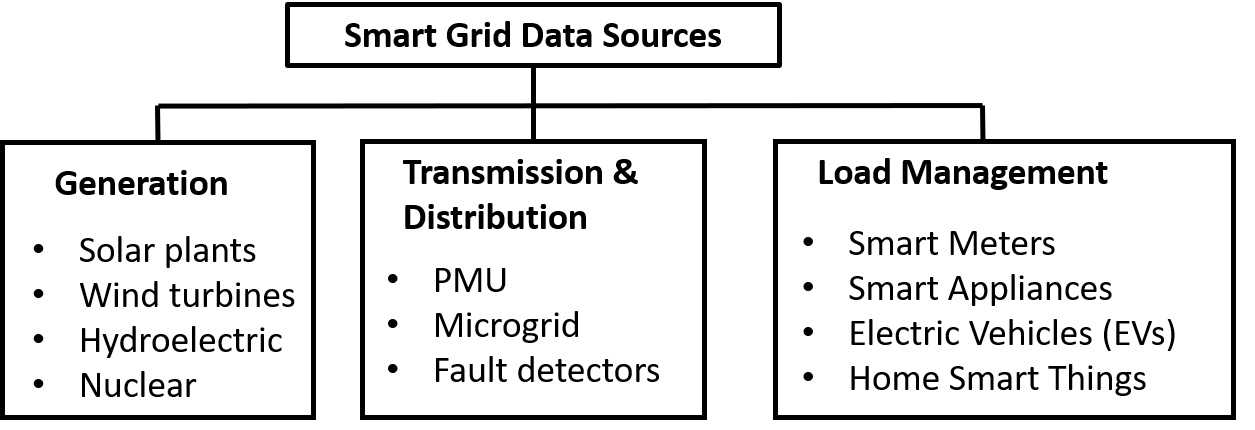}}
\caption{Data Sources in Smart Grid}
\label{fig}
\end{figure}

The cloud frameworks in \cite{b4}\cite{b5} work well when the smart grid data owners own private cloud servers and or the cloud service is completely trusted. However, due to economic reasons, the utility company who owns the smart grid data may choose to outsource the data storage to third-party cloud service providers (\textbf{CSP}). In this scenario, the cloud in the models becomes untrusted, and there is a privacy concern about the smart grid data. Technically speaking, since the data are stored in the CSP's server, the provider can obtain full access to all the sensitive data easily. Moreover, it is reasonable to assume that the CSP may be interested in these data for many reasons. For example, the advanced metering infrastructure (AMI) data in the smart grid can contain customers' personal information, such as hourly measured power consumption, home location, and payment record. By launching side-channel attacks, such as Non-intrusive load monitoring \cite{b24}, it is possible for the CPS to learn the customer's habits and customs, which may bring profit to CPS in many ways. 

One straightforward approach to prevent storage CPS from accessing sensitive data in the smart grid is encrypting the data before uploading them to the cloud. While encryption provides confidentiality to the data, it also sacrifices the functionalities of processing the data. One of the most critical functions of processing data stored in the remote server is searching. For example, a utility company wants to query the billing statement of a specific customer. If the documents are encrypted with the concern of privacy, the CPS can no longer provide the search function to the utility company. In fact, this is a common problem that not only exists in the smart grid field but also in all cloud storage applications that require privacy enhancement.

With the development of privacy-preserving technology, Searchable Symmetric Encryption (SSE) was proposed to address the above problem. SSE is technology that enables users to store documents in ciphertext form while keeping the functionality to search keywords in their documents. In recent years, a series of secure and efficient SSE schemes are proposed \cite{b6}-\cite{b15}. However, most of them are only focusing on general circumstances in which user's documents are collections of random keyword combination, and can become inefficient or over-protect when directly applying them to smart grid applications. 

There are two characteristics that make smart grid data special. Firstly, the smart grid data are believed to be frequently updated and have high generation rate. This also implies there are always new data/keywords generated, which will lead to an increasing keyword dictionary. Furthermore, a large portion of smart grid data are well regulated and have specific structures. In practice, smart grid data may contain multiple attributes that will be searched as keywords. For example, although two utility companies may have different implementations of storing customer billing statement, it is reasonable to assume both implementations should have keywords such as user identity, electricity price, smart meter reading in each record. The two characteristics make the state-of-the-art SSE schemes inappropriate to be applied to smart grid application. The typical SSE schemes together with their disadvantages with the above two characteristics will be discussed in Section II.

In this paper, we present a practical searchable symmetric scheme that aims to protect the privacy of smart grid data. By taking advantage of the two characteristics, our scheme can achieve higher space efficiency with a sacrifice of a little information disclosure which was acceptable for smart grid system in practice. Our main contributions can be summarized as follow:

\begin{itemize}
\item We review and analyze the typical state-of-the-art SSE schemes and show why they are inappropriate for smart grid data.
\item According to the characteristics of smart grid data, we design a simple, practical SSE scheme that provides higher space efficiency with tolerant information leakage in real applications.
\item We implement a prototype based on the statistic data of advanced metering infrastructure (AMI) provided by U.S. Energy Information Administration (EIA) to show the effectiveness of our scheme. We claim the scheme can also be applied to other types of data in the smart grid, such as metering data, customer billing statement, and PMU/PDC data.
\end{itemize}

The rest of paper is organized as follow. Section~\ref{sec:related} gives the introduction of related work on SSE. Section III introduces the threat model and notations we use. The detailed description of our scheme will be presented in Section IV. After that, section V introduces the implementation of the prototype and smart grid data we use. Future work will be discussed in Section VI. Finally, Section VII summarizes the paper.

\section{Related Work} \label{sec:related}

Currently, the fundamental SSE constructions can be classified into three categories, namely construction without indexes, construction with direct index, and construction with inverted indexes. 

The first practical SSE scheme without index construction was proposed by Song \etal in 2000 \cite{b6}. They considered a document as a list of words with the same length and used a specially designed stream cipher to encrypt the document. However, this scheme requires the server to traverse each document word by word, which leads to a search complexity linear to the document size. Furthermore, SSE scheme without index usually requires specially developed encryption algorithms, making it unscalable to current power grid communication and control system. After that, several high impact index-based SSE schemes were proposed. Secure Indexes by Goh \cite{b7} built Bloom Filters as the direct index. By adjusting the parameters of the Bloom Filter, secure indexes can achieve efficient search complexity. However, one inherent problem of Bloom Filter is that it will bring false-positive rate, and this can be unacceptable for critical infrastructure like the smart grid. Another direct index based scheme was presented by Chang \etal \cite{b8}, and they built a large index table for all documents to enable efficient search. However, their scheme assumes that there is a dictionary mapping all keywords to associate identifiers. As discussed in Section I, the number of keywords in smart grid data can be large and keep increasing. Therefore, a scheme in \cite{b8} is also not appropriate for smart grid applications. One of the most famous inverted index based SSE schemes were proposed by Curtmola \etal \cite{b9}. They presented an index scheme which achieves the highest time-efficient search function by using a uniquely designed linked list data structure. However, an inverted index construction scheme has an inherent problem, namely directly updating is difficult. Although a well-designed file management system can mitigate the problem, inverted indexes are still not efficient for smart grid which has new data generated all the time. The latest research work on SSE including dynamic searchable encryption \cite{b10}-\cite{b13}, forward secure searchable encryption \cite{b14}, and fuzzy keyword searchable encryption \cite{b15}. 

Besides the theoretical of SSE scheme, research on applying SSE to solve real-world problems is also active. In general, the SSE scheme can be used to all systems which include storage outsourcing. Tong \etal \cite{b16} employed a modified SSE scheme of Secure Index \cite{b7} and designed a secure data sharing mechanism for situational awareness in the power grid. Their approach is also used to protect the privacy of e-health data \cite{b17}. The problem of health data' privacy is drawing more and more attention. Li \etal \cite{b18} leveraged a secure K-nearest neighbor (KNN) and attribute-based encryption to build a dynamic SSE scheme for e-health data and achieves both forward and backward security. Other applications of SSE can be found in \cite{b19}.

\section{Background} \label{sec:back}

\begin{figure}[htbp]
\centerline{\includegraphics[width=1\linewidth]{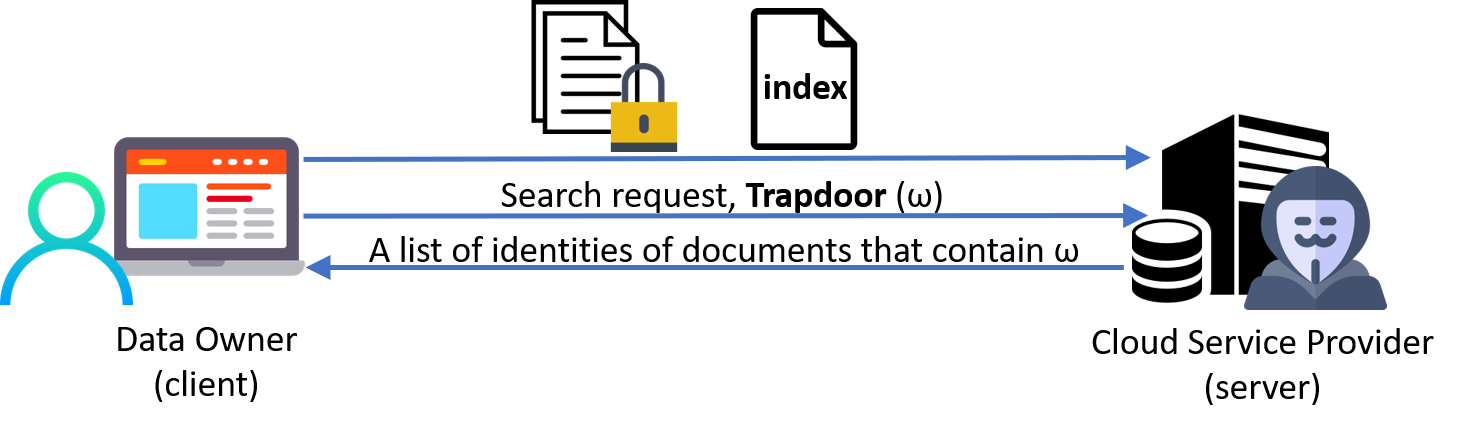}}
\caption{General SSE Working Model} \label{fig:model}
\end{figure}

\subsection{Threat Model and Assumptions}

Our scheme interacts between two parties. As shown in Fig.~\ref{fig:model}, we refer the party who owns the data and wants to outsource the storage of data as the client, and the party who provides the storage service as the server. The client uploads the encrypted data, and associated search index to the server and sends query afterward. For simplicity, our scheme only considers the case that there is one keyword contained in the query. We note that query that contains boolean operation on multiple keywords can be regarded as the operation on query results of multiple single keywords, which will be briefly discussed at the end of this section. After receiving query contains keyword information from the client, the server should run the SSE algorithms and return a list of the identifier of documents that contain the keyword.

Same as most state-of-the-art SSE schemes, in our threat model we also assume that the third party server is a curious but honest attacker, which means the cloud server should provide normal cloud service but will try to learn the content of data. More specifically, the server is not allowed to delete or modify the client's data or share the data with other parties. However, different from SSE schemes which assume that the client's data is a collection of random keyword combination, our scheme makes practical assumptions on the client's data based on the characteristics of smart grid data discussed in Section I. 

First, we assume the smart grid data to be stored in the cloud should be a collection of records that have the same data structure and contain the same number of keywords. For example, the customer billing statement record should have attributes like customer identifier, date, house location, smart meter readings, and additional notes. Another example can be the PMU data. One of the most widely used standards of PMU data is IEEE C37.118.2 \cite{b20}. The standard gives the format of application layer data structure of PMU data that contains several attributes, such as data stream ID number, time stamp, and measurement data. Moreover, as presented in \cite{b16}, the data owner can also design a hierarchical data structure to store the smart grid data. Therefore, we believe this assumption is practical in real power grid applications and is easy to meet. In the rest of the paper, we will use record and document interchangeably for simplicity.

Second, we assume the total number of records to be stored in the cloud is much larger than the number of keywords in each record. In practice, a record may contain up to tens of attributes, but it is normal for a utility company to maintain millions of customer data records or billions of measurement data records. 

One widely used security requirement of searchable symmetric encryption is IND2-CKA secure proposed in \cite{b7}. In brief, IND2-CKA secure requires that the server cannot learn anything more of the plaintext message except for the search result. According to IND2-CKA, the number of keywords in each document should also be kept secret. However, our scheme allows the number of keywords in each record to be known to the attacker (server). There are mainly two practical reasons for this privacy sacrifice. First, SSE schemes follow IND2-CKA secure consider the case that the client stores general documents which contain keywords with random numbers. As we discussed above, since all smart grid data records are assumed to contain the same number of keywords, it is not reasonable to assume that the attacker cannot learn the number in the real working scenario, like by social engineering or just randomly guessing. Second, we will show that our scheme can achieve high space efficiency by loosing the restriction on this privacy leakage issue. This compromise should be acceptable for a utility company or government department in practice.

Finally, the scheme is only used for constructing secure search index. How to protect the privacy of the plaintext data will not be discussed in this paper. The client can use popular security schemes like CBC or CTR block cipher to protect the confidentiality and schemes like HMAC to protect the integrity of data.

\subsection{Notations}

We use $\Delta = \left \{ R_{1}, R_{2} ... R_{2^{d}}\right \}$ to indicate a collection of records. $N$ is defined to be the total number of possible keywords in a record while $n$ is the number of desired keywords in a record that the client want to search, where $n \leq N $. We use $w_{i,j}$ to denote the $jth$ desired keyword in record $R_{j}$ where $1\leq i\leq 2^{d}$ and $1\leq j\leq n$. $ \left \{ 0,1 \right \}^{n}$ is used to present the set of all $n$ bits numbers. $K \overset{R}{\leftarrow} \left \{ 0,1 \right \}^{n}$ means an element $K$ being sampled uniformly from set $ \left \{ 0,1 \right \}^{n}$. In addition, we use $f:\left \{ 0,1 \right \}^{k} \times \left \{ 0,1 \right \}^{r} \rightarrow \left \{ 0,1 \right \}^{r}$ to define a pseudo-random function that maps a $r$ bits number to another $r$ number with a $k$ bits key. A record's identifier is defined as $id(R)$. Finally, we use $h:\left \{ 0,1 \right \}^{*}\rightarrow \left \{ 0,1 \right \}^{r} $ to denote a hash function that maps random length bitwise string to an $r$ bits number.

\section{Practical SSE Scheme Design For Smart Grid} \label{sec:SSE}

\subsection{Construction}

\begingroup
\removelatexerror

\begin{figure*}[htp]
\begin{center}

\fbox{\begin{tabular}{p{0.95\textwidth}}
	
\begin{itemize}

\item $\textbf{Keygen($k$)}$: Given a security parameter $k$, uniformly sample the master key $MK \overset{R}{\leftarrow} \left \{ 0,1 \right \}^{k}$

\item $\textbf{Trapdoor($MK, w$)}$: Given the master key $MK$ and a keyword $w$, generate $T_{w} = f_{MK}(h(w))$ of $w$

\item $\textbf{BuildIndex($R_{i}$, $MK$)}$ : The input is a record $R_{i}$ and the master key $MK$

{\bf{The client:}}

\begin{algorithm}[H]
create an $n$ elements array $I_{R_{i}}$ and initialize all elements to zero\\
create an set $U:\left \{ x\in \mathbb{Z}|0\leq x\leq n-1 \right \}$ \\
\For{each desired keyword $w_{i,j}$ in $R_{i}$}{
compute $T_{w_{i,j}} = $ \textbf{Trapdoor}$(MK, w_{i,j})$  \\
compute $X_{w_{i,j}} = f^{'}_{T_{w_{i,j}}}(h(id(R_{i})))$ \\
pick $\lambda \overset{R}{\leftarrow} U$, update $U = U - \left \{ \lambda \right \}$ \\
set $I_{R_{i}}[\lambda] = X_{i,j}$  \\
							 }
return $(R_{i}, I_{i})$
				\end{algorithm}

\item $\textbf{Search($T_{w}, I_{R}$)}$ : The input is the trapdoor $T_{w}$ of keyword $w$ and the index $I_{R}$ of record $R$

{\bf{The server:}} 

\begin{algorithm}[H]
compute $X_{w} = f^{'}_{T_{w}} (h(id(R)))$\\
\eIf{$X_{w}$ is in list $I_{R}$}{
return 1	}{return 0}
				\end{algorithm}
\end{itemize}
		\end{tabular}
	}

\end{center}
	\setlength{\abovecaptionskip}{-3pt}
	\setlength{\belowcaptionskip}{-10pt}
\caption{Four Polynomial time functions}
\label{basic-construction}
\end{figure*}
\endgroup

Secure Indexes presented in \cite{b7} gives a framework of trapdoor based searchable symmetric encryption scheme which was widely used by the follow-up research. In general, an SSE should consist of four polynomial time algorithms:

\begin{itemize}
\item $\textbf{Keygen($s$)} $ is run by the client to generate a master private key $MK$ where $s$ is a security parameter.
\item $\textbf{Trapdoor($MK$, $w$)}$ is run by the client by taking the master key $MK$ and a keyword $w$ as the input, and outputs the trapdoor $T_{w}$ of word $w$.
\item $\textbf{BuildIndex($R$, $MK$)}$ is run by the client by taking the master key $MK$ and a record $R$ as the input, and outputs the index $I_{R}$ for record $R$. 
\item $\textbf{SearchIndex($T_{w}$, $I_{R}$)}$ is run by the server by taking a trapdoor $T_{w}$ and a document's index $I_{R}$ as the input, and outputs $1$ if $w \in R$ or $0$ otherwise. 
\end{itemize}

In general SSE scheme, the encrypted documents together with the associated indexes will be kept by the server. When searching, the client generates the trapdoor $T_{w}$ for word $w$ and send $T_{w}$ to the server. For each record $R$, the server runs the \textbf{SearchIndex($T_{w}$, $I_{R}$)} function and determine whether $R$ contains $w$. The server will finally return to the client a list of the identifiers of records which contain $w$. The framework requires that only the client who holds the private master key $MK$ can generate the trapdoor $T_{w}$ for each word $w$, such that the server cannot learn related information from the index. Our scheme also follows this framework and is built in the direct index structure.

Our scheme uses a codeword array as the index for a record. For the keyword $w$, the \textbf{Trapdoor} function computes the hash value $h(w)$ of $w$, where $h:\left \{ 0,1 \right \}^{*}\rightarrow \left \{0, 1 \right \}^{r}$, and outputs the trapdoor $T_{w} = f_{MK}(h(w))$, where $f:\left \{ 0,1 \right \}^{k}\times \left \{ 0,1 \right \}^{r}\rightarrow \left \{ 0,1 \right \}^{r}$ is a pseudo-random function. To build the index $I_{R_{i}}$ of record $R_{i}$, the \textbf{BuildIndex} function calls \textbf{Trapdoor} and computes the codeword  $X_{i,j} = f^{'}_{T_{w_{i,j}}}(h(id(R_{i})))$, where $f^{'}:\left \{ 0,1 \right \}^{r}\times \left \{ 0,1 \right \}^{r}\rightarrow \left \{ 0,1 \right \}^{r}$ is another pseudo-random function. Each codeword $X_{i,j}$ should be randomly written into an n element array $I_{R_{i}}$, and this step can be done with a pseudo-random generator in implementation. The detailed design of our scheme can be found in Fig. 3.

As discussed early in this section, we assume the smart grid data $\Delta$ to be a collection of records with each record $R_{i}$ contains $N$ keywords. Considering the scenario that not all types of keywords in the data are necessary for searching, and the data owner only wants to query records by several specific types of keywords. For example, based on the standard IEEE C37.118.2, the utility company may want to search PMU data records by ID number or timestamp, and there may be no need to search by synchronization word. Therefore, our scheme firstly allows the data owner to select the $n \left ( n\leq N \right )$ types of keywords she wants to query based on specific applications. 

After determining $n$ types of keywords, the client should run function $\textbf{Keygen}$ to obtain a $k$ bits master key $MK$ and keep it secret. Subsequently, for each record $R_{i}$ in $\Delta$, the client runs the \textbf{BuildIndex} function to obtain the index $I_{R_{i}}$ of record $R$. Finally, the index $I_{R_{i}}$ should be attached to the encrypted record $R_{i}$ and uploaded to the server. 

To search for a keyword $w$, the client needs to compute the trapdoor $T_{w} = f_{MK}(h(w))$ and sends $T_{w}$ to the server. After receiving the $T_{w}$, the server computes $X_{i,w} = f^{'}_{T_{w}}(h(id(R_{i})))$ for each ciphertext record $R_{i}( 1\leq i\leq 2^{d})$, and checks whether $X_{i,w}$ is contained in $I_{R_{i}}$. If so, the server returns $id(R_{i})$ to the client. 

\IEEEsettopmargin{t}{0.80in}
\begin{table*}[htbp]

\caption{Perfomance Comparison of Various SSE Schemes}
\begin{center}
\begin{tabular}{|c|c|c|c|c|c|c|}
\hline
\textbf{scheme} & \textbf{plaintext encryption}& \textbf{index}& \textbf{false-positive} &\textbf{update} & \textbf{search complexity} & \textbf{index size}\\
\hline
final scheme \cite{b6}& special & no & no & easy & $\mathcal{O}(2^{d}\cdot N)$ & none\\
\hline
Z-IDX \cite{b7} & general & direct & yes & easy &  $\mathcal{O}(2^{d})$ &  $\mathcal{O}(2^{d}\cdot n)$\\
\hline
PPSED \cite{b8} & general & direct & no & easy & $\mathcal{O}(2^{d})$ &  $\mathcal{O}((2^{d})^{2})$\\
\hline
SSE-1 \cite{b9} & general & inverted & no& hard &  $\mathcal{O}(1)$ &  $\mathcal{O}(2^{d}\cdot n)$\\
\hline
our scheme & general & direct & no & easy &  $\mathcal{O}(2^{d}\cdot n)$ &  $\mathcal{O}(2^{d}\cdot n)$\\
\hline
\multicolumn{7}{l}{$2^{d}$is the number of records, $N$ is the number of all keywords in a record. $n$ is the number of desired keywords in a record, where $2^{d} \gg n$.}
\end{tabular}
\end{center}
\end{table*}

\subsection{Analysis and Comparison}

As stated at the beginning of Section III, our scheme aims to protect the data privacy such that the adversary cannot learn any other information about the plaintext record from the index except for the search result and the number of desired keywords in the record. We analyze the security of our scheme from two aspects. First, considering a simple scenario that only one index of a record is given, a polynomial time attacker can not learn the original keywords from the index. This is correct because if the attacker can learn the keyword from the codeword, she will be able to break the pseudo-random function, which is contradictory to the assumptions. Second, we consider the unlinkability of our scheme, which means the attacker is not able to learn whether two records have the same keyword $w$ from their indexes without the trapdoor $T_{w}$. This is achieved by introducing the identifier of records to build the codeword. The same keyword in two records will have different codewords in two indexes. We refer the reader to  \cite{b7} for the mathematical proof.

In general, the main methodology of our scheme is increasing the space efficiency of the SSE scheme with permission of few information leakages based on the characteristics of smart grid data. Our scheme was built based on direct index structure, so it is dynamic and easy-updating. We use the codeword array as the index of each record, which leads to a small index size compared with schemes that involve keyword dictionary. Since the codewords are randomly inserted, the searching complexity of our scheme becomes $\mathcal{O}(2^{d}\cdot n)$. However, since the number of desired keywords $n$ is believed to be a very small constant in practical application, the search algorithm will still be efficient. Our scheme is similar to the PPSED scheme described in \cite{b8}. Both schemes used direct index structure and an array as the index. However, since there is always new keywords (e.g. timestamps) generated in smart grid data, the index size of the PPSED scheme will become extremely large. Table I gives a detailed comparison between our design and the widespread SSE schemes. 

\subsection{Extension}

It is obvious to see that updating new record and associated index to the server is straightforward. The client just needs to run the \textbf{BuildIndex} function of the new record to obtain the index, and appends the (encrypted record, index) pair to the records and indexes stored in the server. 

The plain scheme considers the scenario that only one keyword is searched in a round. In practice, the client may want to search for records that meet specific keyword requirements. One simple example can be a government department who wants to query a dataset of utility companies. A meaningful query can be (State: TN and Establish\_Year: 1998 or 1999). The boolean operations on multiple keywords can be easily applied to the codewords matching process when the server runs \textbf{Search} function.

\section{Implementation} \label{sec:implem}

\subsection{Smart Grid Data}

\begin{table}[htbp]
\caption{Features of AMI Dataset from EIA}
\begin{center}
\begin{tabular}{|c|c|}
\hline
\textbf{category} & \textbf{feature} \\
\hline
Utility 	& Year, Month, Utility Number, \\
Characteristics& Utility Name, State, Data Status\\
\hline
Number AMR - Automated  &  \\
Metering Reading &  \\
\cline{1-1}

Number AMI - Advanced  & Residential, Commercial, \\
Metering Infrastructure & Industrial, Transportation, Total \\
 \cline{1-1}
 
Non AMR/AMI Meters &  \\
 \cline{1-1}
 
Total Number of Meters  &   \\
 \cline{1-1}
 
Energy Saved - AMI (MWh) & \\
 \hline

\end{tabular}
\label{tab2}
\end{center}
\end{table}

The public dataset we used to test the effectiveness of our scheme is the statistical AMI data provided by the U.S. Energy Information Administration \textbf{(EIA)} \cite{b21}. The AMI data are derived from EIA-861M form, which stands for “Monthly Electric Power Industry Report.” The report collects sales of electricity and revenue each month from a statistically chosen sample of electric utilities in the United States. EIA started to collect monthly green pricing, net metering, and advanced metering data since 2011. We choose the CSV file of advanced metering data of the year 2016 as the dataset for our implementation.

As shown in Table II, the CSV contains 31 columns, including year, month, utility name, state, residential AMI, and so on. In our experiment, we select all features except Year from Utility Characteristics and all features from AMI related categories as our desired types of keywords. The CSV file contains 4819 records in total, and the data types include string and unsigned integer.

\subsection{Prototype}

For simplicity, we use Python 2.7 as the programming language for the prototype. We claim that a low-level language like C can be used in practice with the consideration of speed. We build the encryption scheme with the help of Pycrypto \cite{b22}, which is a widely used cryptography library for python. Generally, the ciphers and hash functions in Pycrypto are written in C and provide Python API. We use Advanced Encryption Standard \textbf{(AES)} block cipher with 128 bits key as the pseudo-random function and the MD5 as the hash function described in our scheme. We note that MD5 has been severely compromised and is no longer secure for integrity protection. However, the MD5 in our prototype is only used to generate a unique identifier for record and keyword, similar to the building dictionary process in other schemes.

Since the original ciphertext usually contains unprintable characters that will destroy CSV format, we store the ciphertext as the hexadecimal string to maintain a clear CSV format for demonstration. However, the hexadecimal string will lead to larger index size. Therefore, we suggest that an efficient file-index management system is needed for real-world applications.

The source code is tested on a Mac OS X machine with a 1.6 GHz Intel Core i5 processor and 8GB memory. Our experiment result shows that 4819 indexes can be searched in around 0.15 second. The source code of the prototype is available on Github \cite{b23}.

\section{Future Work}

Our scheme only considers the one client scenario. In practice, it is common for the data owner to grant access permission to a group of people. Attribute-based encryption may be used to address this problem in the future. Meanwhile, same as the PPSED scheme in \cite{b8}, our scheme also suffers forward information leakage, which means the attacker can learn the information of the plaintext from newly added indexes based on previous queries. More information about the SSE forward security can be found in \cite{b14}.

\section{Conclusion} \label{sec:conclu}

In this paper, we firstly review and give a comparison of the state-of-the-art searchable symmetric encryption schemes, and analyze why these schemes are not appropriate for smart grid application. After that, we present a practical scheme based on the characteristics of smart grid data. We show that space efficient and easy update SSE scheme is available by allowing a little information leakage which is acceptable in practice. Finally, we implement a prototype based on statistical AMI data to show the effectiveness of our scheme.

%
%


\begin{thebibliography}{00}

\bibitem{b1} Tan, S., De, D., Song, W.Z., Yang, J. and Das, S.K., 2017. Survey of security advances in smart grid: A data driven approach. IEEE Communications Surveys \& Tutorials, 19(1), pp.397-422.

\bibitem{b2} Arnold, G.W., Wollman, D.A., FitzPatrick, G.J., Prochaska, D.E., Lee, A., Holmberg, D.G., Su, D.H., Hefner Jr, A.R., Golmie, N.T., Simmon, E.D. and Brewer, T.L., 2010. NIST Framework and Roadmap for Smart Grid Interoperability Standards Release 1.0 (No. Special Publication (NIST SP)-1108).

\bibitem{b3} Beres, A., Genge, B. and Kiss, I., 2015. A brief survey on smart grid data analysis in the cloud. Procedia Technology, 19, pp.858-865.

\bibitem{b4} Arenas-Martinez, M., Herrero-Lopez, S., Sanchez, A., Williams, J.R., Roth, P., Hofmann, P. and Zeier, A., 2010, October. A comparative study of data storage and processing architectures for the smart grid. In Smart Grid Communications (SmartGridComm), 2010 First IEEE International Conference on (pp. 285-290). IEEE.

\bibitem{b5} Rusitschka, S., Eger, K. and Gerdes, C., 2010, October. Smart grid data cloud: A model for utilizing cloud computing in the smart grid domain. In Smart Grid Communications (SmartGridComm), 2010 First IEEE International Conference on (pp. 483-488). IEEE.

\bibitem{b6} Song, D.X., Wagner, D. and Perrig, A., 2000. Practical techniques for searches on encrypted data. In Security and Privacy, 2000. S\&P 2000. Proceedings. 2000 IEEE Symposium on (pp. 44-55). IEEE.

\bibitem{b7} Goh, E.J., 2003. Secure indexes. IACR Cryptology ePrint Archive, 2003, p.216.

\bibitem{b8} Chang, Y.C. and Mitzenmacher, M., 2005, June. Privacy preserving keyword searches on remote encrypted data. In International Conference on Applied Cryptography and Network Security (pp. 442-455). Springer, Berlin, Heidelberg.

\bibitem{b9} Curtmola, R., Garay, J., Kamara, S. and Ostrovsky, R., 2011. Searchable symmetric encryption: improved definitions and efficient constructions. Journal of Computer Security, 19(5), pp.895-934.

\bibitem{b10} Kamara, S., Papamanthou, C. and Roeder, T., 2012, October. Dynamic searchable symmetric encryption. In Proceedings of the 2012 ACM conference on Computer and communications security (pp. 965-976). ACM.

\bibitem{b11} Miers, I. and Mohassel, P., 2016. IO-DSSE: Scaling Dynamic Searchable Encryption to Millions of Indexes By Improving Locality. IACR Cryptology ePrint Archive, 2016, p.830.

\bibitem{b12} Lu, Y., 2012, February. Privacy-preserving Logarithmic-time Search on Encrypted Data in Cloud. In NDSS.

\bibitem{b13} Hahn, F. and Kerschbaum, F., 2014, November. Searchable encryption with secure and efficient updates. In Proceedings of the 2014 ACM SIGSAC Conference on Computer and Communications Security (pp. 310-320). ACM.

\bibitem{b14} Bost, R., 2016, October. $\Sigma$o$\varphi$o$\zeta$: Forward Secure Searchable Encryption. In Proceedings of the 2016 ACM SIGSAC Conference on Computer and Communications Security (pp. 1143-1154). ACM.

\bibitem{b15} Li, J., Wang, Q., Wang, C., Cao, N., Ren, K. and Lou, W., 2010, March. Fuzzy keyword search over encrypted data in cloud computing. In INFOCOM, 2010 Proceedings IEEE (pp. 1-5). IEEE.

\bibitem{b16} Tong, Y., Deyton, J., Sun, J. and Li, F., 2013. $ S^{3} A $: A Secure Data Sharing Mechanism for Situational Awareness in The Power Grid. IEEE Transactions on Smart Grid, 4(4), pp.1751-1759.

\bibitem{b17} Tong, Y., Sun, J., Chow, S.S. and Li, P., 2014. Cloud-assisted mobile-access of health data with privacy and auditability. IEEE Journal of biomedical and health Informatics, 18(2), pp.419-429.

\bibitem{b18} Li, H., Yang, Y., Dai, Y., Bai, J., Yu, S. and Xiang, Y., 2017. Achieving Secure and Efficient Dynamic Searchable Symmetric Encryption over Medical Cloud Data. IEEE Transactions on Cloud Computing.

\bibitem{b19} Bösch, C., Hartel, P., Jonker, W. and Peter, A., 2015. A survey of provably secure searchable encryption. ACM Computing Surveys (CSUR), 47(2), p.18.

\bibitem{b20} IEEE Standard for Synchrophasor Data Transfer for Power Systems," in IEEE Std C37.118.2-2011 (Revision of IEEE Std C37.118-2005) , vol., no., pp.1-53, Dec. 28 2011
doi: 10.1109/IEEESTD.2011.6111222

\bibitem{b21} www.eia.gov/electricity/data/eia861m/index.html\#ammeter

\bibitem{b22} www.dlitz.net/software/pycrypto/

\bibitem{b23} github.com/jiangnan3/Searchable\_Encryption

\bibitem{b24} Hart, George William. "Nonintrusive appliance load monitoring." Proceedings of the IEEE 80.12 (1992): 1870-1891.


\end{thebibliography}
\end{document}